\documentclass[
twocolumn,
preprintnumbers,
nofootinbib,
prd,aps
]{revtex4}

\usepackage{epsf}
\usepackage{latexsym}
\usepackage{amssymb}
\usepackage{epsf}
\usepackage{amsmath}
\usepackage{slashed}

\begin{document}


\newcommand{\rem}[1]{{\bf #1}}

\renewcommand{\topfraction}{0.8}

\preprint{UT-11-40}

\title{

Imprints of Cosmic Phase Transition in Inflationary Gravitational Waves

}

\author{
Ryusuke Jinno, Takeo Moroi and Kazunori Nakayama
}

\affiliation{
Department of Physics, University of Tokyo, Bunkyo-ku, 
Tokyo 113-0033, Japan
}

\date{December, 2011}

\begin{abstract}

  We discuss the effects of cosmic phase transition on the spectrum of
  primordial gravitational waves generated during inflation.  The
  energy density of the scalar condensation responsible for the phase
  transition may become sizable at the epoch of phase transition,
  which significantly affects the evolution of the universe.  As a
  result, the amplitudes of the gravitational waves at high frequency
  modes are suppressed.  Thus the gravitational wave spectrum can be a
  probe of phase transition in the early universe.

\end{abstract}

\maketitle

\renewcommand{\thefootnote}{\#\arabic{footnote}}

Spontaneous symmetry breaking (SSB) often plays very important role in
high energy physics.  For the construction of the standard model of
particle physics, which is currently the most successful model of high
energy phenomena, the SSB of $SU(2)_L\times U(1)_Y\rightarrow
U(1)_{\rm em}$ due to the Higgs mechanism (i.e., the electroweak
symmetry breaking) is crucial.  In addition, at the scale of the
quantum chromodynamics (QCD), chiral symmetry breaking also occurs.
If we consider various models of physics beyond the standard model,
SSBs may occur at higher energy scales.  For example, if the strong CP
problem is solved by the Peccei-Quinn (PQ) mechanism
\cite{Peccei:1977hh}, the PQ symmetry breaking should happen at the PQ
scale.  In grand unified theories (GUTs) \cite{Georgi:1974sy}, the
symmetry breaking of $G_{\rm GUT}\rightarrow SU(3)_C\times
SU(2)_L\times U(1)_Y$ occurs at the GUT scale (where $G_{\rm GUT}$ is
the GUT gauge group).  In supersymmetric models, the supersymmetry
breaking terms are expected to arise due to the spontaneous breaking
of supersymmetry.

The SSBs in the framework of the standard model, i.e., the electroweak
symmetry breaking and the chiral symmetry breaking in QCD, may be well
understood in the future by experimental data (in particular, by the
LHC result), lattice simulation, and so on.  However, it is difficult
to study the SSBs in models beyond the standard model because their
energy scales are too high to be reached by collider experiments.
Thus, the physics related to those SSBs at high energy scales are
hardly probed by the existing methods.

If we consider cosmology, there should exist in the past a period of
cosmic phase transition related to the SSB.  In particular, in a
period around the phase transition, the expansion of the universe may
be significantly affected by the energy density of the fields which
cause the SSB, which results in a significant deviation from the
radiation-dominated universe.  In the following, we will show that
information of such an early universe may be imprinted in the spectrum
of primordial gravitational waves (GWs) generated during inflation.
This is because the GW spectrum is sensitive to the expansion history
of the universe~\cite{Seto:2003kc, Smith:2005mm, Boyle:2005se,
  Nakayama:2008ip, Kuroyanagi:2008ye, Nakayama:2009ce,
  Kuroyanagi:2011fy}.  Importantly, the GW spectrum may be studied at
future experiments such as DECIGO~\cite{Seto:2001qf} and/or
BBO~\cite{gr-qc/0506015}.

In this letter, we discuss the possibility of studying cosmic phase
transition in the early universe by using GW.  We will show that the
evolution of the amplitudes of GW may be significantly affected if
phase transition happened in the early epoch and hence the information
on the SSB may be extracted from the spectrum of GW.

Let us first discuss how the amplitudes of GWs evolve in the expanding
universe.  The tensor perturbation of the metric, which corresponds to
the degrees of freedom of the GW, is defined as
\begin{eqnarray}
  ds^2 = - dt^2 + a^2 (t) \left( \delta_{ij} + 2h_{ij} \right) dx^i dx^j,
\end{eqnarray}
where $ds$ is the line element, and the indices $i$ and $j$ run $1-3$.
The tensor perturbation $h_{ij}$ satisfies transverse and traceless
conditions, $h_i^i={h_{ij}}^{,j}=0$.  Thus, there are two physical
degrees of freedom (for a fixed value of the momentum), which we
denote by $+$ and $\times$. In the following discussion, it is
convenient to work in the momentum space, so we define the Fourier
amplitude of $h_{ij}$ as
\begin{eqnarray}
  h_{ij}(t,\vec{x}) = \frac{1}{M_{\rm Pl}}
  \sum_{\lambda=+,\times} \int \frac{d^3 \vec{k}}{(2\pi)^3}
  \tilde{h}^{(\lambda)}_{\vec{k}} (t)
  \epsilon_{ij}^{(\lambda)} e^{i\vec{k}\vec{x}},
\end{eqnarray}
where $M_{\rm Pl}\simeq 2.4\times 10^{18}\ {\rm GeV}$ is the reduced
Planck scale, and $\epsilon_{ij}^{(\lambda)}$ is the polarization
tensor which satisfies
$\epsilon_{ij}^{(\lambda)}\epsilon_{ij}^{(\lambda')}=\delta_{\lambda\lambda'}$.
Then, $\tilde{h}^{(\lambda)}_{\vec{k}}$ satisfies
\begin{eqnarray}
  \ddot{\tilde{h}}^{(\lambda)}_{\vec{k}} + 
  3 H \dot{\tilde{h}}^{(\lambda)}_{\vec{k}} +
  \frac{k^2}{a^2(t)} \tilde{h}^{(\lambda)}_{\vec{k}} = 0,
  \label{waveeq}
\end{eqnarray}
where the ``dot'' denotes the derivative with respect to time, and
$k\equiv|\vec{k}|$.  Here, we neglect the anisotropic stress which is
irrelevant for the present study.  Although we will numerically follow
the evolution of the GW amplitudes, it is instructive to shortly
discuss the qualitative behavior of the solution of Eq.\
\eqref{waveeq}.  When $k\ll aH$ (out of horizon), the last term of the
left-hand side is irrelevant and hence $\tilde{h}_{\vec{k}}$ stays
(almost) constant.  On the contrary, once the mode enters the horizon
(i.e., $k\gg aH$), $\tilde{h}_{\vec{k}}$ is under oscillation; in this
case, $\langle \tilde{h}^2_{\vec{k}}\rangle_{\rm osc}$ and $\langle
\dot{\tilde{h}}^2_{\vec{k}}\rangle_{\rm osc}$ are approximately
proportional to $a^{-2}$ and $a^{-4}$, respectively, where
$\langle\cdots\rangle_{\rm osc}$ denotes the average for the time
scale much longer than the oscillation time (but shorter than cosmic
time).

During inflation, the quantum fluctuation of the GW determines the
initial value of $\tilde{h}^{(\lambda)}_{\vec{k}}$.  
The present GW spectrum per log wavenumber interval for $k_{\rm EW}\ll
k \ll k_{\rm RH}$ (where $k_{\rm EW}$ and $k_{\rm RH}$ are comoving
wavenumbers of modes which enter the horizon at the time of
electroweak phase transition and reheating after inflation,
respectively) is given by
\begin{eqnarray}
  \Omega_{\rm GW}^{\rm (SM)} (k) \simeq 1.7\times 10^{-15} r_{0.1} \gamma
  ~:~
  k_{\rm EW}\ll k \ll k_{\rm RH}.
\end{eqnarray}
Here, $r_{0.1}$ is the tensor-to-scalar ratio in units of $0.1$ and
\begin{equation}
  \gamma = \left[\frac{g_*(T_{\rm in}(k))}{g_{*0}} \right]
  \left[\frac{g_{*s0}}{g_{*s}(T_{\rm in}(k))} \right]^{4/3}
  \left( \frac{k}{k_0} \right)^{n_t},
\end{equation}
where $T_{\rm in}(k)$ denotes the temperature at which the mode $k$
enters the horizon, $g_*$ and $g_{*s}$ denote the effective number of
relativistic degrees of freedom for the energy density and the entropy
density, respectively, with subscript $0$ being for the present value.
In addition, $k_0=0.002{\rm Mpc}^{-1}$ is the pivot scale and $n_t$ is
the tensor spectral index, which is given by $n_t=-r/8$ in standard
inflation models.  Thus the scale dependence of the primordial GW
spectrum is very weak as long as $r$ is small enough.  Hereafter, we
neglect the scale dependence for simplicity.

Now let us discuss how the universe expands in the period of cosmic
phase transition.  During the cosmic phase transition, the expectation
value of the order parameter changes from zero to a finite value due
to thermal effects. The detail of the phase transition depends on the
physics in the SSB sector.  In the present study, we model the SSB
sector simply by introducing a scalar field $\phi$ which plays the
role of the order parameter.  With $\phi$ and $\chi$ being (real)
scalar fields, we consider the following scalar potential,
\begin{eqnarray}
  V (\phi) = \frac{1}{4!} g (\phi^2 - v_\phi^2)^2 
  + \frac{1}{2} h \chi^2 \phi^2,
\end{eqnarray}
where $g$ and $h$ are coupling constants while $v_\phi$ is the vacuum
expectation value of $\phi$.\footnote
{ Problematic domain wall formations are avoided if one regards $\phi$
  as a radial part (absolute value) of the complex scalar field, which
  triggers the SSB of a continuous symmetry, such as U(1).  Cosmic
  strings in association with the SSB of U(1) are not harmful for
  $v_\phi \lesssim 10^{15}$\,GeV. }
Here, $\phi$ represents the scalar field responsible for the cosmic
phase transition; $\phi=0$ ($\phi=v_\phi$) corresponds to the
symmetric (broken) phase, while $\chi$ represents the degrees of
freedom in thermal bath (with temperature $T$).

With $\chi$ being in thermal bath, the free energy of $\phi$ acquires
a term proportional to $T^2\phi^2$ around $\phi =0$.  Thus, in the
early universe with high enough temperature, the symmetry is expected
to be restored.  As the temperature decreases, the negative
mass-squared term in the (zero-temperature) scalar potential wins over
the thermal mass term, and the SSB takes place.  Here, we
expect that the evolution of $\phi$ is well governed by the following
equation,
\begin{eqnarray}
  \ddot{\phi} + 3 H \dot{\phi} + V_T' = - \Gamma_\phi \dot{\phi},
\end{eqnarray}
where $H$ is the expansion rate of the universe, $\Gamma_\phi$ is the
decay rate of $\phi$, and the ``dash'' denotes derivative with respect
to $\phi$.  In addition, $V_T$ is the potential of $\phi$ in thermal
bath, which we evaluate as $V_T=\frac{1}{4} g (\phi^2 - v_\phi^2)^2 +
\frac{1}{2} h \langle \chi^2 \rangle_T \phi^2$, where $\langle{\cal
  O}\rangle_T\equiv{\rm tr}[{\cal O}\rho]$ (with $\rho$ being density
matrix) is thermal average of the operator ${\cal O}$.  We consider
the case where the time scale of the change of $\phi$ is much smaller
than that of cosmic expansion.  Thus, we take $\rho =e^{-H_\chi/T}$,
where $H_\chi$ is the Hamiltonian in $\chi$-sector to calculate
$\langle\chi^2\rangle_T$.

We evaluate the thermal average by approximating $\chi$ as a free
scalar field with the mass squared of $h\phi^2$ and obtain
\begin{eqnarray}
  V_T (\phi) &=& \frac{1}{4!} g (\phi^2 - v_\phi^2)^2 
  \nonumber \\ &&
  + \frac{1}{2} h \phi^2
  \int \frac{p^2 dp}{2\pi^2 \omega_{p,\phi}}
  \frac{1}{e^{\omega_{p,\phi}/T} - 1},
  \label{V_T}
\end{eqnarray}
where $\omega_{p,\phi}=\sqrt{p^2 + h\phi^2}$.\footnote
{One may estimate the expectation value of $\phi$ by minimizing the
  free energy.  In the present analysis, we approximate that the phase
  transition occurs when the curvature of the potential at $\phi$
  becomes zero.  With such an approximation, results of free-energy
  and our procedures agree because the thermal mass of the scalar
  field $\phi$ obtained in two procedures are the same.}
For the following discussion, it is instructive to expand the
potential around $\phi =0$ to find
\begin{eqnarray}
  V_T (\phi) = \frac{1}{24} h (T^2 - T_{\rm c}^2) \phi^2 + \cdots,
\end{eqnarray}
where
\begin{eqnarray}
  T_{\rm c} = \sqrt{ \frac{2g}{h} } v_\phi.
\end{eqnarray}
Thus, the curvature at $\phi =0$ changes its sign at $T=T_{\rm c}$.  

Because we are interested in the case where $\Gamma_\phi\gg H$, the
oscillation of $\phi$ decays away with the time scale much faster than
the cosmic expansion.  Thus, we can approximate that $\phi$ follows
the temporal minimum of $V_T$.  Even so, an accurate understanding of
the evolution of $\phi$ is not straightforward because, in some
period, $V_T$ has two minima.  In particular, at the temperature just
above $T_{\rm c}$, $\phi =0$ is not the absolute minimum of the
potential, and hence the phase transition may proceed with a tunneling
from $\phi =0$ to the absolute minimum~\cite{Callan:1977pt}.  Whether
the phase transition completes via the first or second order phase
transition depends on the parameters in the model. For $h\ll 1$, the
second order phase transition precedes the first order one.  For
$h\sim O(1)$, the first order phase transition may take place at the
temperature $T\sim ({\rm a~few})\times T_{\rm c}$.  In either case, we
simply approximate that the phase transition occurs at the time when
the cosmic temperature becomes $T_{\rm c}$ for the first time.  In
addition, because the position of the true minimum of the potential at
$T=T_{\rm c}$ well agrees with that at $T=0$, we approximate the
energy density of the $\phi$ sector as
\begin{eqnarray}
  \rho_\phi = \left\{ \begin{array}{ll}
      V(0) & ~:~ t<t_{\rm PT} \\
      0 & ~:~ t>t_{\rm PT} \\
    \end{array} \right. ,
\end{eqnarray}
where $t_{\rm PT}$ is the time of the phase transition.

With the above approximation, the evolution of the scale factor $a$ is
governed by
\begin{eqnarray}
  H^2 \equiv
  \left( \frac{\dot{a}}{a} \right)^2 = 
  \frac{\rho_{\rm rad} + \rho_\phi}{3 M_{\rm Pl}^2},
\end{eqnarray}
where $\rho_{\rm rad}$ is the energy density of the radiation
component; it evolves as
\begin{eqnarray}
  \dot{\rho}_{\rm rad} + 4 H \rho_{\rm rad} = 
  V(0) \delta (t-t_{\rm PT}),
\end{eqnarray}
and is related to the cosmic temperature as $\rho_{\rm rad} =
\frac{\pi^2}{30} g_* T^4$, with $g_*$ being the effective number of
massless degrees of freedom.  In our analysis, we take the
standard-model value of $g_*$, which is $106.75$.
Although $\chi$ may also contribute to $g_*$,
such a contribution is so small that we can safely neglect it.

\begin{figure}[t]
  \begin{center}
    \centerline{\epsfxsize=0.45\textwidth\epsfbox{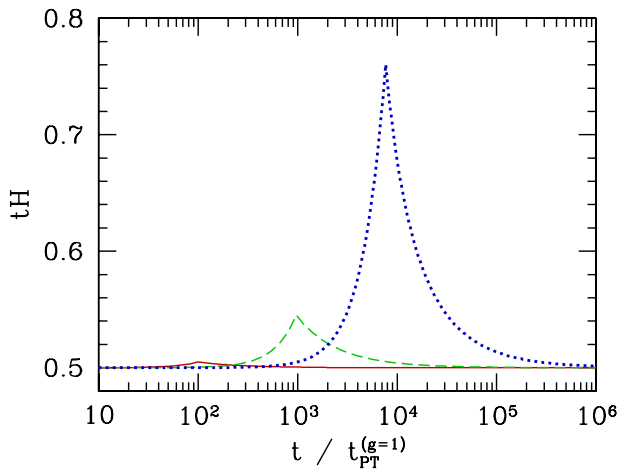}}
    \caption{Evolution of the product $tH$ as a function of time
      (normalized by $t_{\rm PT}^{(g=1)}$) for $g=10^{-2}$
      (red-solid), $10^{-3}$ (green-dashed), and $10^{-4}$
      (blue-dotted).  Here we take $h=1$.  Notice that the figure is
      independent of $v_\phi$.}
    \label{fig:bg}
  \end{center}
\end{figure}

In Fig.\ \ref{fig:bg}, we plot the product of the time $t$ and the
expansion rate $H$ as a function of $t$ for several values of $g$.
(In the plot, $t$ is normalized by $t_{\rm PT}^{(g=1)}$, which is the
time of the phase transition for the case of $g=1$, to make the figure
independent of $v_\phi$.)  The product $tH$ is equal to $\frac{1}{2}$
if the universe is dominated by radiation.  We can see that the
evolution of the universe at $t\sim t_{\rm PT}$ deviates from that of
radiation-dominated universe as $g$ becomes smaller.  This behavior
can be easily understood from the relation $[\rho_\phi/\rho_{\rm
  rad}]_{t=t_{\rm PT}}\sim O(h^2/g_* g)$.  For smaller $g$, the
potential energy of $\phi$ at the origin tends to dominate the
universe before the phase transition.  In the small $g$ limit, a brief
period of inflation takes place \cite{Lyth:1995ka}.

\begin{figure}[t]
  \begin{center}
   \centerline{\epsfxsize=0.45\textwidth\epsfbox{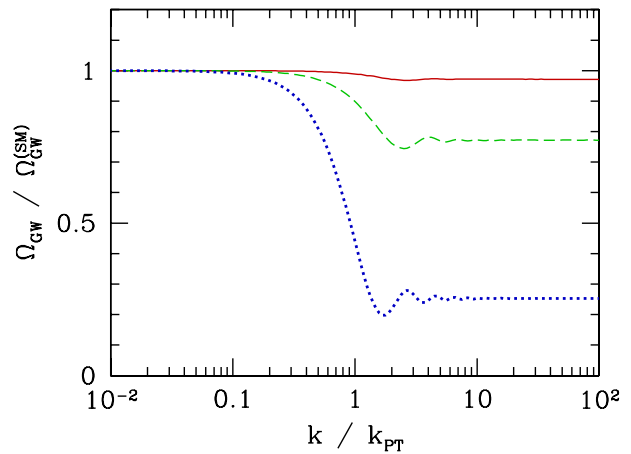}}
   \caption{$\Omega_{\rm GW}(k)/\Omega_{\rm GW}^{\rm SM}(k)$ as a
     function of $k$ (normalized by $k_{\rm PT}$) for $g=10^{-2}$
     (red-solid), $10^{-3}$ (green-dashed), and $10^{-4}$
     (blue-dotted).  Here we take $h=1$.  Notice that the figure is
     independent of $v_\phi$.}
    \label{fig:gw}
  \end{center}
\end{figure}

Once the evolution of the scale factor is understood, we can easily
solve Eq.\ \eqref{waveeq} to obtain the present spectrum of the GW.
In Fig.\ \ref{fig:gw}, we plot the present GW spectrum as a function
of $k$.  Assuming that $T_{\rm c}$ is much higher than the electroweak
scale, we normalize $\Omega_{\rm GW}$ as
\begin{eqnarray}
  \Omega_{\rm GW} (k\ll k_{\rm PT}) = 
  \Omega_{\rm GW}^{\rm (SM)} (k),
\end{eqnarray}
where $k_{\rm PT}\equiv a_{\rm PT}H_{\rm PT}$ is the wavenumber of the
mode which enters the horizon at the time of the phase transition.

One can see that the GW spectrum with $k\gtrsim k_{\rm PT}$ is
suppressed.  With the present approximation, the following relation
holds,
\begin{eqnarray}
  R \equiv
  \left.
    \frac{\Omega_{\rm GW} (k)}{\Omega_{\rm GW}^{\rm (SM)} (k)}
  \right|_{k\gg k_{\rm PT}}
  = \frac{\rho_{\rm rad}(T_{\rm c})}{\rho_{\rm rad}(T_{\rm c})+V(0)},
  \label{R_PT}
\end{eqnarray}
where $R$ is the reduction rate of the high-frequency GW spectrum due
to the phase transition.  The right-hand side of Eq.\ \eqref{R_PT}
depends only on the combination of $g/h^2$, and is independent of
$v_\phi$.  For $g/h^2=1\times 10^{-4}$ ($3\times 10^{-4}$, $1\times
10^{-3}$, $3\times 10^{-3}$, $1\times 10^{-2}$), $R$ is given by
$0.25$ ($0.50$, $0.76$, $0.91$, $0.97$).  If a short period of
inflation occurs with sufficiently small $g/h^2$, the spectrum of GWs
which enter the horizon during such a period is proportional to
$k^{-4}$.

In order to discuss the possibility of studying the cosmic phase
transition using GWs, it is necessary to understand the present
frequency of the mode with $k\sim k_{\rm PT}$.  (The comoving
wavenumber is related to the present frequency as $f=k/2\pi a_0$, with
$a_0$ being the present scale factor.)  Let us define
\begin{eqnarray}
  T_{\rm PT} \equiv
  \left( \frac{4g^2}{h^2} + \frac{5g}{4\pi^2 g_*} \right)^{1/4}
  v_\phi,
\end{eqnarray}
which is the temperature just after the phase transition.
Then, the present frequency of the mode with $k=k_{\rm PT}$ is given by
\begin{eqnarray}
  f_{\rm PT} \simeq 
  2.7\ {\rm Hz} \times 
  \left( \frac{T_{\rm PT}}{10^8\ {\rm GeV}} \right).
\end{eqnarray}
We have seen that $g/h^2\ll 1$ when $1-R$ becomes sizable.  In such a
case, we obtain
\begin{eqnarray}
  \left[ f_{\rm PT} \right]_{g/h^2\ll 1} \simeq
  0.50\ {\rm Hz} \times 
  \left( \frac{g^{1/4} v_\phi}{10^{8}\ {\rm GeV}} \right).  \label{fPT}
\end{eqnarray}

Finally we discuss the possibility for detecting characteristic
features of the phase transition in the GW spectrum.  For this
purpose, we approximate the sensitivity of the future space
interferometers such as DECIGO/BBO with correlation analysis of 1 year
\cite{Seto:2001qf, gr-qc/0506015, gr-qc/0511145, gr-qc/9909001} as
\begin{equation}
  \Omega_{\rm sens}(f)=\left\{
    \begin{array}{ll}
      \Omega^{\rm (min)}_{\rm sens}(f_1/f) & ~:~ f<f_1,\\
      \Omega^{\rm (min)}_{\rm sens}(f/f_1)^3 & ~:~ f_1<f<f_2,\\
      \Omega^{\rm (min)}_{\rm sens}(f_2/f_1)^3(f/f_2)^5 & ~:~ f_2<f,\\
    \end{array}
  \right.
\end{equation}
with $\Omega^{\rm (min)}_{\rm sens}=3\times 10^{-18}$, $f_1=0.3\ {\rm
  Hz}$, and $f_2=2\ {\rm Hz}$.  Then, we expect that the modulation in
the GW spectrum due to the phase transition is in the range of
detector sensitivity if (i) $(1-R)\Omega_{\rm GW}(f_{\rm PT}) >
\Omega_{\rm sens}(f_{\rm PT})$, and (ii) $R\Omega_{\rm GW}(f_{\rm PT})
> \Omega_{\rm sens}({\rm max}\{ f_{\rm PT}, f_1\})$.  Condition (i)
ensures that the drop-off of $\Omega_{\rm GW}$ is larger than the
sensitivity, while condition (ii) means that the GWs with $k\gtrsim
k_{\rm PT}$ are observable.  For $(r,f_{\rm PT})=(0.1,0.1\ {\rm Hz})$,
$(0.1,1\ {\rm Hz})$, and $(0.01,0.1\ {\rm Hz})$, for example, the
conditions (i) and (ii) are satisfied when $0.005<R<0.98$,
$0.17<R<0.83$, and $0.05<R<0.86$, respectively.  Broader regions will
be explored in the ultimate-DECIGO \cite{Seto:2001qf,gr-qc/0511145},
where sensitivities will be improved by orders of magnitude.  Notice
that, because of the stochastic background from white dwarf binaries,
it will be difficult to extract the signal of cosmic phase transition
in the GW spectrum if $f_{\rm PT}\lesssim 0.1\ {\rm Hz}$
\cite{astro-ph/0304393}.

As a final remark, the scalar field dynamics associated with phase
transitions produces GWs of flat spectrum \cite{Krauss:1991qu,
  Fenu:2009qf, Giblin:2011yh}.  This contribution is small enough to
be neglected for the intermediate scale phase transition with $v_\phi
\sim 10^8\,$GeV, which we are interested in (see Eq.\ (\ref{fPT})).

In summary, we have argued that the spectrum of GW can be a useful
probe of the cosmic phase transition in the early universe.  So, if
GWs with sizable amplitudes (i.e., the sizable value of the
tensor-to-scalar ratio parameter $r$) are observed by the measurement
of $B$-mode polarization of the cosmic microwave background, we have a
good chance of studying the cosmic phase transition in the early
universe with precise observations of primordial GWs.

{\it Acknowledgment}: This work is supported by Grant-in-Aid for
Scientific research from the Ministry of Education, Science, Sports,
and Culture (MEXT), Japan, No.\ 22540263 (T.M.), No.\ 22244021 (T.M.),
No.\ 23104001 (T.M.), No.\ 21111006 (K.N.), and No.\ 22244030 (K.N.).


\end{document}